\documentclass[reprint,aps,pra,showpacs,
superscriptaddress,longbibliography]{revtex4-1}
\usepackage{graphicx}
\usepackage{amsmath}
\usepackage{amssymb}
\usepackage{epstopdf}
\usepackage{amsfonts}
\usepackage{dcolumn}
\usepackage{bm}
\usepackage{color}
\newcolumntype{d}[1]{D{.}{.}{#1}}

\begin{document}

\title{Trapped unitary two-component Fermi gases with up to ten particles}
\author{X. Y. Yin}
\affiliation{Department of Physics and Astronomy, Washington State University, Pullman,
Washington 99164-2814, USA}
\author{D. Blume}
\affiliation{Department of Physics and Astronomy, Washington State University, Pullman,
Washington 99164-2814, USA}

\date{\today }

\begin{abstract}
The properties of two-component Fermi gases
with zero-range interactions are universal.
We use an explicitly correlated Gaussian basis set expansion
approach to investigate small
equal-mass two-component Fermi gases under spherically
symmetric external harmonic confinement. At unitarity,
we determine the ground state energy for systems 
with up to ten particles interacting
through finite-range two-body potentials for
both even and odd number of particles.
We extrapolate the energies to the zero-range limit
using a novel scheme that removes the linear
and, in some cases, also the quadratic dependence of the ground
state energies on the two-body range.
Our extrapolated zero-range energies are compared with
results from the literature.
We also calculate the two-body
Tan contact and structural properties.
\end{abstract}

\maketitle

\section{Introduction}
\label{sec_intro}
The properties of two-component Fermi gases 
interacting through two-body zero-range potentials with
$s$-wave scattering length $a_s$ are 
universal~\cite{giorginirmp,blumerpp}.
At unitarity, i.e., for infinitely large $a_s$,
the two-body interaction does not define a meaningful
length scale and the strongly-interacting
Fermi gas is characterized by the same number of
length scales as the non-interacting Fermi gas.
Approximate realizations of the unitary
Fermi gas include 
dilute neutron matter in the crusts of
neutron stars~\cite{carlson08} 
and ultracold atom gases
such as ${}^6$Li~\cite{hulet01}
and ${}^{40}$K~\cite{jin99}. 
The properties of homogeneous and
inhomogeneous unitary Fermi gases have
attracted a great deal of experimental and
theoretical attention. 
For spherically symmetric external confinement,
the harmonic oscillator length $a_\text{ho}$ defines the only
length scale of the system. It is hence interesting to determine
how the properties of trapped unitary Fermi gases
vary with the number of particles. 

Harmonically trapped Fermi gases at unitarity
have been treated by quantum Monte Carlo 
methods~\cite{bertsch07, blume07, blume08, kaplan10, 
kaplan11, forbes12, carlson14,  alhassid13}, 
density functional theory (DFT)~\cite{bulgac07, forbes12, 
salasnich08, adhikari09a, adhikari09b},
and basis set expansion approaches.
The accuracy of the fixed-node
diffusion Monte Carlo (FN-DMC) energies~\cite{blume07,
blume08, forbes12, carlson14}
depends on the quality
of the many-body nodal surface. The resulting
energies provide upper bounds to the exact ground
state energies and the
zero-range limit is reached 
through extrapolation~\cite{forbes12, carlson14}. 
Auxiliary-field quantum Monte
Carlo (AFMC) methods, on the other hand, work on a finite lattice
and extrapolation to the infinite
lattice limit is required to obtain fully converged 
result~\cite{carlson14}. 
The quality of DFT
calculations depends
critically on the underlying functional. Since
the functional is typically obtained by matching
to data for the homogeneous system, the
analysis of results for the trapped system
can provide insights into
gradient corrections and other finite size 
features~\cite{salasnich08,  adhikari09a, adhikari09b}.
Trapped unitary Fermi gases with up to six
particles have been calculated by explicitly correlated
Gaussian (ECG) basis set expansion approaches~\cite{cg_rmp, 
blume07, blume08, blume09, javier09, blume10,
debraj12} with better
than about $1\%$ accuracy.
Application of the ECG method to systems with
more than six particles has been challenging due to
the rapid increase
of the number of permutations and the larger number
of degrees of freedom.
Recently, Ref.~\cite{quiney14}
treated the $(N_1, N_2)=(4,4)$ system at unitarity
using a basis set that accounts for the
most important but not all correlations.

Here,
we present results for small trapped unitary
Fermi gases with $N \leq 10$ particles, where $N=N_1+N_2$
and $N_1-N_2=0$ or $1$.
Our extrapolated zero-range energy of the (4,4)
system
is $0.9\%$ lower than that reported
in Ref.~\cite{quiney14}.
A new aspect of our work is that we developed an improved
scheme for extrapolating the finite-range energies to
the zero-range limit. This new scheme eliminates the
linear and, in some cases, the quadratic dependence of the ground
state energies on the two-body range.
The scheme provides a consistency
check on the range-dependence of our energies and
reduces the errors that result from the extrapolation
to the zero-range limit.
Our results suggest that the developed range correction
scheme allows one to obtain a reliable approximation
to the zero-range energy from a single finite-range
calculation. The scheme can be applied to 
other numerical calculations that work with finite-range 
interactions. We use our range correction scheme to
determine the zero-range energies and the Tan contact for
two-component Fermi gases with $N\leq10$ at unitarity.
In addition, we
present selected structural properties.

The remainder of this paper is organized as follows.
Section \ref{sec_theory} discusses the theoretical
framework and our extrapolation scheme
to the zero-range limit.
Section \ref{sec_result} presents our results for
systems with up to ten particles and compares,
where available, with results from the literature. 
Lastly, Sec. \ref{sec_conclusion}
concludes.

\section{Theoretical framework}
\label{sec_theory}
We consider equal-mass two-component Fermi gases
with $N_1$ spin-up and $N_2$ spin-down atoms
($N=N_1+N_2$ and $N_1-N_2=0$ or $1$) 
under external spherically symmetric
harmonic confinement with angular trapping frequency
$\omega$. The system Hamiltonian $H(r_0)$ reads
\begin{eqnarray}
\label{eq_H}
H(r_0)=\sum_{i=1}^{N} -\frac{\hbar^2}{2m}
\nabla_{i}^2
+V_\text{tr}(\vec{r}_1, ..., \vec{r}_N) \nonumber \\
+
\sum_{i=1}^{N_1}\sum_{j=N_1+1}^{N}
V_{2b}(r_{ij}, r_0),
\end{eqnarray}
where $m$ denotes the atom mass,
$\vec{r}_i$ denotes
the position vector of the $i$th particle with respect to the
trap center, and
\begin{eqnarray}
V_\text{tr}(\vec{r}_1, ..., \vec{r}_N)
=\sum_{i=1}^{N} \frac{1}{2}m\omega^2\vec{r}_i^2
\end{eqnarray}
is the trapping potential.
$V_{2b}$ is the interspecies two-body interaction
potential that depends on the interparticle distance 
$r_{ij}$, $r_{ij}=|\vec{r}_i-\vec{r}_j|$.
In our work, it is modeled
by a finite-range Gaussian potential with range $r_0$ and
depth $V_0$ ($V_0<0$),
\begin{eqnarray}
V_{2b}(r, r_0)=V_0 \exp \bigg( -\frac{r^2}{2r_0^2} \bigg).
\end{eqnarray}
For a fixed $r_0$, $V_0$ is adjusted such that
$V_{2b}(r, r_0)$ has
an infinitely large $s$-wave scattering length $a_s$
and supports one zero-energy two-body bound state 
in free space.
The ranges $r_0$ considered depend on
the size of the system and vary from $0.01a_\text{ho}$ 
to $0.12a_\text{ho}$, where
$a_\text{ho}$ denotes the harmonic oscillator
length [$a_\text{ho}=\sqrt{\hbar/(m\omega)}$].
For the Gaussian potential with one zero-energy bound state,
the effective range $r_\text{eff}$
is approximately equal to $2.032r_0$.

To numerically solve the Schr\"odinger equation for the
Hamiltonian given in Eq.~(\ref{eq_H}),
we separate off the center of mass degrees of freedom and
expand the eigenstates of the relative Hamiltonian
in terms of ECG basis functions, which depend on a set of non-linear
variational parameters that are optimized through
energy minimization (see below)~\cite{cg_book, cg_rmp}. 
The unsymmetrized basis functions for states with
$L^{\pi}=0^{+}$ and $1^{-}$ symmetry 
($L$ denotes the relative orbital angular momentum
and $\pi$ the relative parity)
read $\exp(-\frac{1}{2}\vec{x}^{T}A\vec{x})$ and
$\mathcal{Y}_{10}(\vec{u}^{T}\vec{x})
\exp(-\frac{1}{2}\vec{x}^{T}A\vec{x})$, respectively,
where $A$ is a symmetric and positive definite
$(N-1)\times(N-1)$ parameter matrix,
$\vec{u}=(u_1, u_2, ..., u_{N-1})^T$ is a 
$N-1$ dimensional vector, and $\mathcal{Y}_{10}$
is a solid spherical harmonic function~\cite{cg_book}.
$\vec{x}=(\vec{x}_1, \vec{x}_2, ..., \vec{x}_{N-1})^T$ 
collectively denotes a set of
$N-1$ Jacobi vectors.
The ground state of even $N$ systems has $0^{+}$
symmetry and that of odd $N$ systems has $1^{-}$
symmetry.
A key advantage of these basis functions is that the
corresponding overlap and Hamiltonian matrix
elements can be calculated analytically~\cite{cg_book}.

The fermionic exchange symmetry is ensured by acting
with the permutation operator $\mathcal{A}$ on the
unsymmetrized basis functions. 
The number of permutations $N_p$ increases factorially with
the number of identical fermions. For the $(5,5)$ system, e.g.,
$\mathcal{A}$ contains $(5!)^2=14,400$ two-particle
exchange operations with alternating plus and minus signs.
The evaluation of each overlap and Hamiltonian
matrix element involves a sum over $N_p$ terms
that are highly oscillatory. 
In a standard 16 digit floating point implementation,
numerical challenges arise from
the near-cancellation of the positive and negative terms
for systems with $N>8$. The near-cancellation
of these terms of alternating signs can be interpreted as
a relative of the fermion sign problem known from 
Monte Carlo simulations~\cite{mc_book, sign}. 
To ensure that the matrix
elements for the largest systems
considered are accurate to at least ten significant digits,
we implemented our C codes using extended precision. 
The eigenenergies and expansion coefficients are obtained
by solving a generalized eigenvalue problem
that involves the Hamiltonian matrix and the overlap matrix.
The numerical error of the resulting eigenenergies
is several orders of magnitude smaller than the errors
that arise from the use of a finite basis set and the extrapolation
to the zero-range limit.
In Sec.~\ref{sec_result}, we report the total ground state
energy $E(r_0)$ of the Hamiltonian $H(r_0)$,
i.e., we add the center of mass energy of $3E_\text{ho}/2$
to the relative energy obtained by the ECG approach.
Here, $E_\text{ho}$ denotes the harmonic oscillator energy
($E_\text{ho}=\hbar\omega$).

\begin{figure}[htbp]
\includegraphics[angle=0,width=56mm]{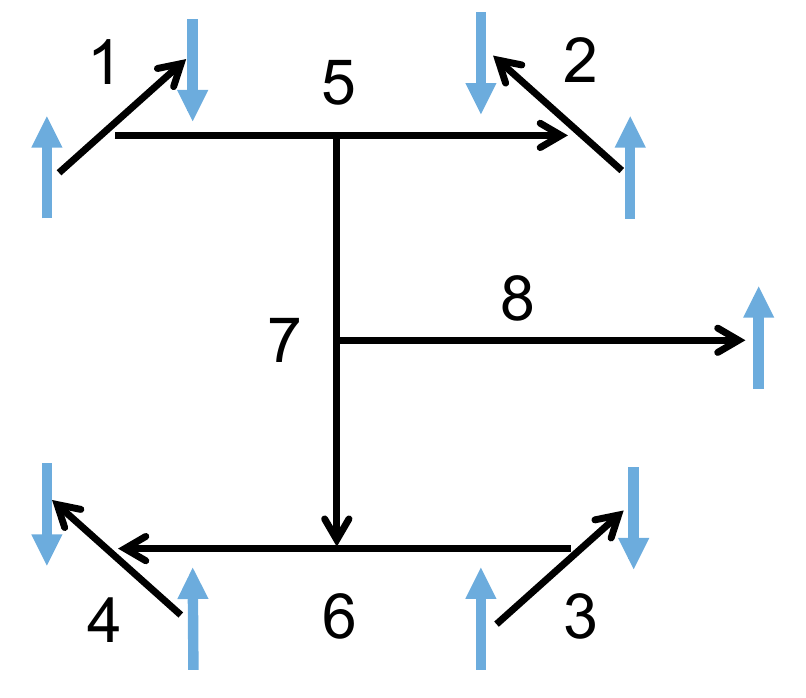}
\caption{(Color online) Illustration of the
Jacobi coordinates employed in our
work for the $(N_1,N_2)=(5,4)$ system.
The dark vectors show the Jacobi vectors
$\vec{x}_1$, $\vec{x}_2$, ..., $\vec{x}_8$.
The spin-up and spin-down fermions are represented
by light vertical up and down arrows.}
\label{fig_jacobi}
\end{figure}

We use a semi-stochastic 
variational approach to choose
and optimize the variational parameters contained in
$A$ and $\vec{u}$~\cite{cg_book}. Our Jacobi coordinates are
chosen such that the first $N_2$ Jacobi vectors
correspond to distance vectors between
unlike particle pairs.
The next $N_2/2$ Jacobi vectors correspond to
the distance vectors between the center of mass
of the first pair and the second pair,
the distance vector between the center of mass
of the third pair and the fourth pair, and so on.
The remaining Jacobi vectors connect the
larger sub-units and, for odd $N$, the $N$th
particle (see Fig.~\ref{fig_jacobi} for an
illustration for $N=9$).
For this choice of Jacobi coordinates, 
the first $N_2$ diagonal elements
of the $A$ matrix represent correlations between unlike particles.
We expect that these interspecies
distances are, on average, smaller and more strongly
correlated than those between like particles.
This motivates us to choose the 
first $N_2$ diagonal elements
of the $A$ matrix for each basis function
from a preset non-linear grid. 
The other diagonal elements are chosen
stochastically from preset parameter windows. 
As in Ref.~\cite{quiney14},
we start with basis functions
that are diagonal in $A$. These basis functions account for
the most important correlations.
Once a certain basis set size is reached (for
the larger systems around $N_b=500$),
we reoptimize the
variational parameters contained
in the diagonal of $A$, and for odd $N$ in $\vec{u}$,
and allow for
off-diagonal $A$ matrix elements.
The off-diagonal matrix element at
position $(i, j)$ of the $A$ matrix is chosen
from $0$ to the geometric mean of the 
$i$th and $j$th diagonal elements.
We find that this choice results
with high probability in positive-definite $A$ matrices.
The positive-definiteness
of the $A$ matrix is tested through diagonalization.
We refer to the reoptimization of all variational
parameters contained in $A$ and, if applicable,
$\vec{u}$ of all $N_b$ basis functions as a reoptimization cycle. 
The re-optimization cycle is repeated until
the lowest energy changes by less than a preset value.
After that, we extend the basis set by a 
hundred to several hundred basis functions and reoptimize the
variational parameters of the enlarged basis set using a
variable number of reoptimization cycles.
This process is repeated a few times. At the end,
the basis is enlarged to around
2000 basis functions without additional
reoptimization of the non-linear variational parameters.
The basis set errors reported in
Tables~\ref{table32} and \ref{table44} of
Sec.~\ref{sec_result}
and Tables I-VI of the Supplemental Material~\cite{sm}
are estimated by analyzing the energy decrease that
results from the basis set enlargements and the reoptimization
cycles.

To reach the universal regime where $a_\text{ho}$ defines the
only length scale in the system, we need to extrapolate
the numerically calculated finite-range energies to the
zero-range limit. 
In previous ECG works~\cite{cg_rmp, blume09, blume10,
debraj12, quiney14}, this was
done by fitting the finite-range energies by a linear or
quadratic function. We refer to this traditional
extrapolation scheme as the zeroth-order extrapolation scheme.
The  difference between the finite-range energies
and the extrapolated zero-range energies is, typically, at the order 
of a few percent and can introduce a non-negligible extrapolation
error.
Moreover, for larger systems, it is computationally
expensive, maybe even prohibitively expensive,
to obtain energies at very small ranges.
It should also be noted that
the extrapolated zero-range energies do not provide
variational upper bounds even though the finite-range ECG
energies do.
It is thus desirable to remove the linear and, ideally, quadratic
range dependence.
Motivated by the generalized virial theorem
\begin{eqnarray}
E(0)=2V_\text{tr}(0)
\end{eqnarray} 
[$V_\text{tr}(0)$ denotes the expectation value
of $V_\text{tr}(\vec{r}_1, ..., \vec{r}_N)$ for 
$r_0\rightarrow0$]
at unitarity,
Werner~\cite{werner08} proposed to remove 
the linear range-dependence of the ground state energy
by combining it with the expectation value $V_\text{tr}(r_0)$
of the trapping potential $V_\text{tr}(\vec{r}_1, ..., \vec{r}_N)$
calculated for the same $r_0$,
\begin{eqnarray}
\label{eq_virial}
E(0)=3E(r_0)-4V_\text{tr}(r_0)+\mathcal{O}(r_0^2).
\end{eqnarray}
While Eq.~(\ref{eq_virial}) removes the leading-order
range dependence, it is associated with errorbars that
come from the basis set errors of 
$E(r_0)$ and $V_\text{tr}(r_0)$.
In our ECG method, the basis set is optimized by
minimizing the ground state energy. Not surprisingly, 
we find that the
convergence of the expectation value of the trapping potential
is not as good as that of the energy.
This motivates us to 
propose an alternative scheme that can be carried out
to higher orders.

The ground state energy $E(r_0)$ of the $N$-particle system is
a smooth function of the two-body interaction range
$r_0$.
The $n_\text{max}$th order Taylor series of
$E(\bar{r}_0)$ around $r_0$ is
\begin{eqnarray}
\label{eq_taylor1}
E(\bar{r}_0)=\sum_{n=0}^{n_\text{max}}
E^{(n)}(r_0)\frac{1}{n!}(\bar{r}_0-r_0)^n \\ \nonumber
+\mathcal{O}[(\bar{r}_0-r_0)^{n_\text{max}+1}],
\end{eqnarray}
where
\begin{eqnarray}
\label{eq_derivative}
E^{(n)}(r_0)=\frac{\partial^n E(\bar{r}_0) }{\partial \bar{r}_0^n}
\bigg|_{\bar{r}_0=r_0}
\end{eqnarray}
is the $n$th order derivative of the ground state
energy with respect to the range evaluated at $r_0$.
$E^{(0)}(r_0)$ is simply the ground state energy
$E(r_0)$
of $H(r_0)$.
$E^{(1)}(r_0)$ can be obtained
through the Hellmann-Feynman theorem~\cite{feynman},
\begin{eqnarray}
E^{(1)}(r_0)
=\left\langle 
\frac{\partial H(\bar{r}_0) }{\partial \bar{r}_0} \bigg|_{\bar{r}_0=r_0}
\right\rangle,
\end{eqnarray}
which is exact in the limit that the basis
set is complete.
The matrix elements needed to evaluate $E^{(1)}(r_0)$ 
reduce to compact analytical expressions.
$E^{(2)}(r_0)$ can be obtained by
the finite difference method, i.e., by
evaluating $E^{(1)}(r_0)$ at two nearby
$r_0$.

Our goal is to obtain the zero-range
energy $E(0)$. Setting $\bar{r}_0$ in Eq.~(\ref{eq_taylor1})
to $0$, 
we obtain
\begin{eqnarray}
\label{eq_taylor2}
E(0)=E_{\text{ZRA}, n_\text{max}}(r_0)
+\mathcal{O}(r_0^{n_\text{max}+1}),
\end{eqnarray}
where
\begin{eqnarray}
\label{eq_taylor3}
E_{\text{ZRA}, n_\text{max}}(r_0)=
\sum_{n=0}^{n_\text{max}}
E^{(n)}(r_0)\frac{1}{n!}(-r_0)^n
\end{eqnarray}
is the $n_\text{max}$th order
approximation to the zero-range energy.
Equations~(\ref{eq_taylor2}) and (\ref{eq_taylor3}) 
establish
a relation between the zero-range energy $E(0)$ and
the finite-range energy $E(r_0)$
and its derivatives with respect to the two-body range.
$E_{\text{ZRA}, 0}(r_0)$ is simply the finite-range
energy $E(r_0)$ with linear leading-order
range-dependence. 
The leading-order range-dependence of
$E_{\text{ZRA}, 1}(r_0)$ and
$E_{\text{ZRA}, 2}(r_0)$
is quadratic and cubic, respectively.
Crucial is that $E_{\text{ZRA}, 1}(r_0)$ and
$E_{\text{ZRA}, 2}(r_0)$
are obtained at finite $r_0$ 
without extrapolation. They
provide better approximations to the zero-range
energy $E(0)$ than $E_{\text{ZRA}, 0}(r_0)$.
We refer to the extrapolations of
$E_{\text{ZRA}, 1}(r_0)$ and
$E_{\text{ZRA}, 2}(r_0)$ 
to the zero-range limit
as the first- and second-order extrapolation schemes. 
For a complete basis, $E_{\text{ZRA},1}(r_0)$
coincides with the quantity
$3E(r_0)-4V_\text{tr}(r_0)$, i.e., formally Eq.~(\ref{eq_taylor2})
with $n_\text{max}=1$ is equivalent to Eq.~(\ref{eq_virial}).
It turns out, however, that our ECG implementation
provides a more accurate estimate for
$E^{(1)}(r_0)$ than for $V_\text{tr}(r_0)$.
In Sec.~\ref{sec_result}, we independently
fit $E_{\text{ZRA}, 0}(r_0)$, 
$E_{\text{ZRA}, 1}(r_0)$, and
$E_{\text{ZRA}, 2}(r_0)$
and compare the resulting zero-range energies.
Appendix  A shows that the functional forms of
$E_{\text{ZRA}, 0}(r_0)$, 
$E_{\text{ZRA}, 1}(r_0)$, and
$E_{\text{ZRA}, 2}(r_0)$ are correlated
and presents the results of a single combined
fit of $E_{\text{ZRA}, 0}(r_0)$, 
$E_{\text{ZRA}, 1}(r_0)$, and
$E_{\text{ZRA}, 2}(r_0)$.
The resulting zero-range energy is found to
be consistent with the zero-range energies
obtained from the independent fits.

The ECG calculations 
become numerically more challenging
with decreasing two-body interaction range $r_0$. 
The challenges arise from the need to resolve
length scales of different orders of magnitude.
In previous ECG calculations~\cite{blume10, debraj12}, 
much effort was put
on solving the Schr\"odinger equation for
systems with small $r_0$.
In this work, we show that a reliable approximation to
the zero-range energy can be obtained
by calculating
$E_{\text{ZRA}, 2}(r_0)$ at
a range $r_0\approx0.1a_\text{ho}$.
We demonstrate in Sec.~\ref{sec_result} that 
$E_{\text{ZRA}, 1}(r_0)$ and
$E_{\text{ZRA}, 2}(r_0)$ play important
roles in obtaining the zero-range energy $E(0)$ and
its errorbar.

To calculate the two-body Tan contact $C(r_0)$ at
unitarity, we use the adiabatic
energy relation~\cite{tan08},
\begin{eqnarray}
\label{eq_tan1}
C(r_0)=\frac{4\pi m}{\hbar^2} 
\frac{\partial E(r_0)}{\partial(-a_s^{-1})}
\bigg|_{a_s^{-1}=0}.
\end{eqnarray}
To obtain the two-body contact for
$r_0=0$, we use the zeroth- and
first-order zero-range extrapolation schemes, i.e.,
we extrapolate $C_{\text{ZRA},0}(r_0)$ and 
\begin{eqnarray}
\label{eq_tan2}
C_{\text{ZRA}, 1}(r_0)=C(r_0)
-
\frac{\partial C(\bar{r}_0)}{\partial \bar{r}_0}
\bigg|_{\bar{r}_0\rightarrow r_0}
r_0
\end{eqnarray}
to the zero-range limit.

The contact can alternatively be calculated through
the pair relation
\begin{eqnarray}
\label{eq_tan3}
C(r_0)=N_1\times N_2\times\lim_{r\rightarrow 0,r\gg r_0}
(4\pi)^2 P_{12}(r)r,
\end{eqnarray}
where $P_{12}(r)$ denotes the pair distribution function.
The quantity $r^2P_{12}(r)$
 with
normalization 
$4\pi \int_{0}^{\infty} P_{12}(r) r^2 dr=1$
tells one the likelihood of finding two unlike particles
at distance $r$ from each other.
The behavior of $4\pi P_\text{12}(r)r^2$
around $r\approx r_0$ depends on the
details of the two-body interaction potential.
Specifically, for finite-range potentials
the quantity $4\pi P_\text{12}(r)r^2$ goes
to zero as $r\rightarrow0$. For the
zero-range potential, in contrast, $4\pi P_\text{12}(r)r^2$
remains finite as $r\rightarrow0$. 
Thus, to extract the finite-range contact via the pair relation,
we consider the region where $r\gg r_0$ but
$r\ll a_\text{ho}$. In practice, the condition
$r\gg r_0$ translates to $r\gtrsim 2r_0$

We also consider
the spherically symmetric radial density $P_{j}(r)$
of species $j$, $j=1$ and $2$.
For even $N$, we have $P_1(r)=P_2(r)$.
The quantity  $P_{j}(r)$
tells one the likelihood
of finding a particle
at distance $r$ from the trap center.
The normalization is chosen such that
$4\pi \int_{0}^{\infty} P_{j}(r) r^2 dr=1$.

\section{Results}
\label{sec_result}

\begin{table*}[htbp]
\caption{Ground state
energy of the (3,2) system at unitarity.
Column 2 shows the finite-range energy
for the largest basis set considered.
The estimated basis set error $\Delta E(r_0)$ 
is reported in column 3.
Columns 4 and 5 report the quantities
$E^{(1)}(r_0)$ and $E^{(2)}(r_0)$;
errorbars are given in parenthesis. 
The energy derivatives are calculated for the largest
basis set considered.
Columns 6-8 report the energies 
$E_{\text{ZRA}, 0}(r_0)$,
$E_{\text{ZRA}, 1}(r_0)$, 
and $E_{\text{ZRA}, 2}(r_0)$.
These energies account for the estimated
basis set extrapolation error, i.e.,
$E(r_0)-\Delta E(r_0)$ is being used to
calculate $E_{\text{ZRA}, j}(r_0)$
for $j=0$, $1$, and $2$.
The errorbars of 
$E_{\text{ZRA}, 1}(r_0)$
and $E_{\text{ZRA}, 2}(r_0)$
account for the uncertainties of
$E^{(1)}(r_0)$ and $E^{(2)}(r_0)$
but do not account for the uncertainty
of $\Delta E(r_0)$.
The last row reports
the extrapolation of $E_{\text{ZRA}, j}(r_0)$
to the zero-range limit.}
\centering
\begin{tabular}{c c c c c c c c}
\hline
\hline
$\frac{r_0}{a_\text{ho}}$ 
&  $\frac{E(r_0)}{E_\text{ho}}$ & 
$\frac{\Delta E(r_0)}{E_\text{ho}} $ &
$E^{(1)}(r_0) \frac{a_\text{ho}}{E_\text{ho}}$
& 
$E^{(2)}(r_0)
\frac{a_\text{ho}^2}{E_\text{ho}}$
& $\frac{E_{\text{ZRA}, 0}(r_0)}{E_\text{ho}}$
& $\frac{E_{\text{ZRA}, 1}(r_0)}{E_\text{ho}}$
& $\frac{E_{\text{ZRA}, 2}(r_0)}{E_\text{ho}}$
\\
\hline
0.07 & 7.5449 &  0.0001
&1.148(7) & $-5.13(12)$ & 7.5448 & 7.4644(5) & 7.4518(8)\\
0.06 & 7.5332 & 0.0001
&1.201(8) & $-4.61(25)$ & 7.5331 & 7.4610(5) & 7.4527(9)\\
0.05 & 7.5211 & 0.0004
&1.240(12) & $-4.11(28)$ & 7.5207 & 7.4587(6) & 7.4536(10)\\
0.04 & 7.5084 & 0.0004
&1.284(15) & $-2.85(33)$ & 7.5080 & 7.4566(6) & 7.4543(9)\\
0.03 & 7.4954 & 0.0006
&1.297(26) & $-0.32(46)$ & 7.4948 & 7.4559(8) & 7.4557(10)\\
0.02 & 7.4825 & 0.0009
&1.290(120) & 0.24(96) & 7.4816 & 7.4558(24) & 7.4559(26)\\
\hline
0 & & & & & 7.4557 &  7.4550 & 7.4563 \\
\hline
\hline
\end{tabular}
\label{table32}
\end{table*}

\begin{figure}[htbp]
\includegraphics[angle=0,width=56mm]{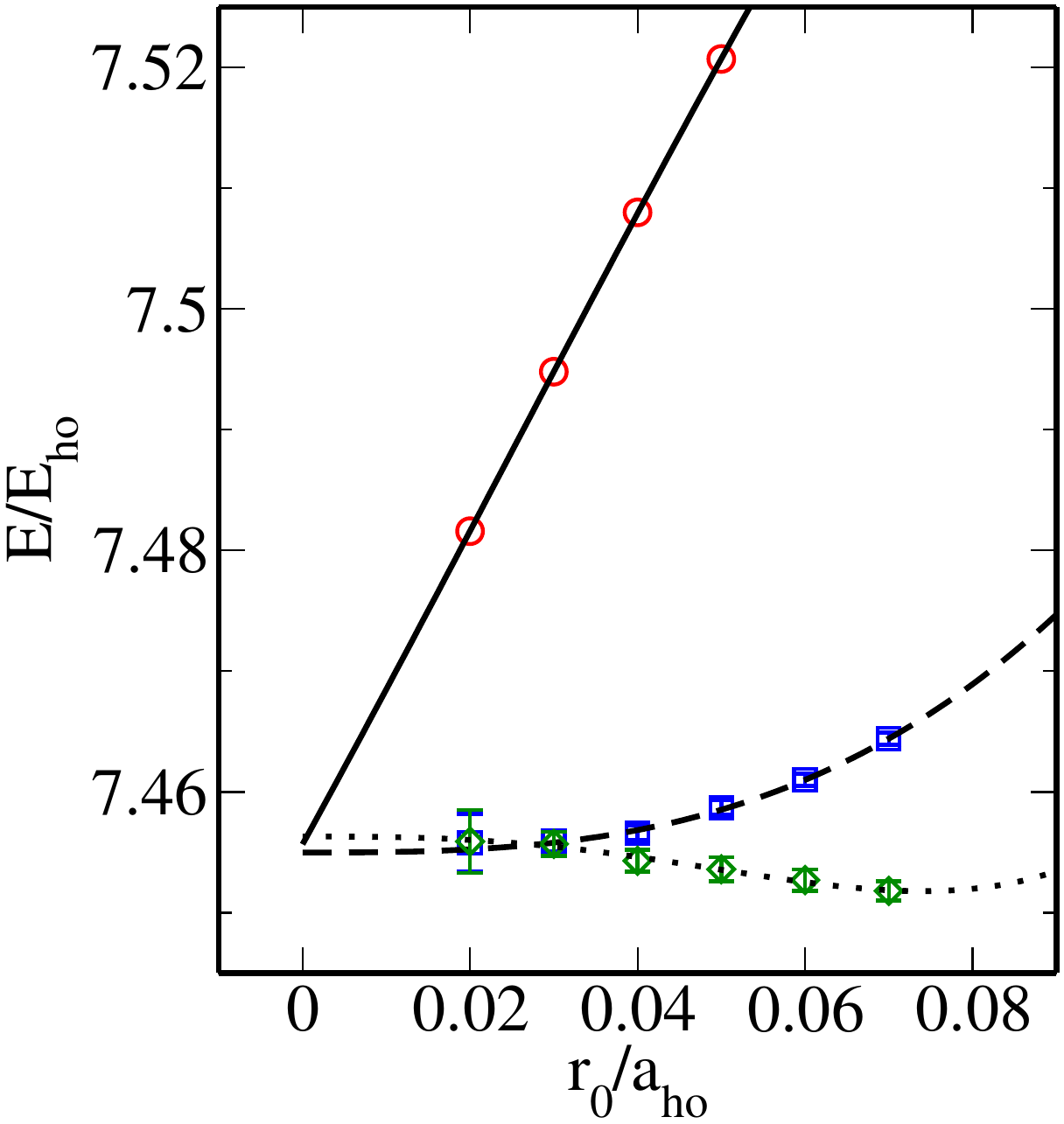}
\caption{(Color online) Ground state energy
of the (3,2) system at unitarity as a function
of $r_0$. 
Circles, squares, and diamonds show the energies
$E_{\text{ZRA},0}(r_0)$, $E_{\text{ZRA}, 1}(r_0)$, 
and $E_{\text{ZRA}, 2}(r_0)$,
respectively, reported in the last three columns of
Table~\ref{table32}.
The uncertainty of $\Delta E(r_0)$
is not accounted for by the errorbars.
Solid, dashed, and dotted lines show polynomial fits
to $E_{\text{ZRA},0}(r_0)$, $E_{\text{ZRA}, 1}(r_0)$, 
and $E_{\text{ZRA}, 2}(r_0)$.}
\label{fig_energy32}
\end{figure}

This section discusses the energies and other
observables of the systems with $N\leq10$
obtained by the ECG method.
We use the (3,2) system to explain our new
range correction scheme.
Column 2 of Table~\ref{table32}
shows the finite-range
energies $E(r_0)$ obtained by the ECG approach
for various two-body interaction ranges $r_0$.
The reported energies are obtained for
the largest basis set considered.
They provide variational upper bounds for the finite-range
Hamiltonian with Gaussian interaction.
Column 3 reports the estimated
basis set error
$\Delta E(r_0)$.
For all $r_0$ considered, the basis set error
is less than $0.02\%$.
Columns 4 and 5 show the quantities
$E^{(1)}(r_0)$ and $E^{(2)}(r_0)$, respectively.
While $E^{(1)}(r_0)$ increases slightly with decreasing
range $r_0$, this increase is smaller than the decrease of $r_0$,
implying that the range correction $E^{(1)}(r_0)r_0$
decreases with decreasing
$r_0$.
The magnitude of $E^{(2)}(r_0)$ decreases with decreasing
$r_0$. Note that we are not able to
estimate $E^{(2)}(r_0)$ reliably for small $r_0$
(the errorbars are larger than the quantity itself).
Yet, the errorbars of $E^{(2)}(r_0)$ allow us to
estimate the maximal correction proportional to $r_0^2$
for each $r_0$, thereby providing us
with another means to estimate errorbars.
Columns 6-8 of Table~\ref{table32}
show 
$E_{\text{ZRA}, j}(r_0)$ 
with $j=0$, $1$, and $2$. These
values are obtained by subtracting the basis
set error $\Delta E(r_0)$.
The leading-order range
dependence of $E_{\text{ZRA}, 0}(r_0)$ is linear and
we perform a fit of the form
$c_0+c_1 r_0+c_2 r_0^2+c_3 r_0^3$.
The extrapolated zero-range energy
is reported in the last row of column 6.
The leading order range dependence of
$E_{\text{ZRA}, 1}(r_0)$ is quadratic and
we perform a fit of the form
$c_0+c_2 r_0^2+c_3 r_0^3+c_4 r_0^4$,
weighted by the inverse square of 
the uncertainty. 
The extrapolated zero-range energy 
is reported
in the last row of column 7.
The leading-order range dependence of
$E_{\text{ZRA}, 2}(r_0)$ is cubic and
we perform a fit of the form
$c_0+c_3 r_0^3+c_4 r_0^4$,
weighted by the inverse square of 
the uncertainty.
The extrapolated zero-range energy
is reported
in the last row of column 8. 
Table~\ref{table32}
shows that the zeroth-, first- and second-order
extrapolation schemes yield zero-range energies
that differ by at most $0.0013E_\text{ho}$.
This confirms that the range-dependence of the
(3,2) ground state energy for the $r_0$ considered is
well described by a Taylor series.
Moreover, we note that $E_{\text{ZRA}, 2}(r_0)$
for
$r_0=0.07a_\text{ho}$ differs by only $0.0045E_\text{ho}$
from the extrapolated zero-range energy.
This suggests that $E_{\text{ZRA}, 2}(r_0)$
obtained at a single (relatively large) range
provides
a very good estimate for the zero-range energy.
Circles, squares, and diamonds in Fig.~\ref{fig_energy32}
show the energies $E_{\text{ZRA}, 0}(r_0)$,
$E_{\text{ZRA}, 1}(r_0)$, and $E_{\text{ZRA}, 2}(r_0)$, respectively.
The fits (see the discussion above) are shown by lines.

\begin{table*}[htbp]
\caption{Same as Table~\ref{table32} but for
the (4,4) system at unitarity.
}
\centering
\begin{tabular}{c c c c c c c c}
\hline
\hline
$\frac{r_0}{a_\text{ho}}$ 
&  $\frac{E(r_0)}{E_\text{ho}}$ & 
$\frac{\Delta E(r_0)}{E_\text{ho}} $ &
$E^{(1)}(r_0) \frac{a_\text{ho}}{E_\text{ho}}$
& 
$E^{(2)}(r_0)
\frac{a_\text{ho}^2}{E_\text{ho}}$
& $\frac{E_{\text{ZRA}, 0}(r_0)}{E_\text{ho}}$
& $\frac{E_{\text{ZRA}, 1}(r_0)}{E_\text{ho}}$
& $\frac{E_{\text{ZRA}, 2}(r_0)}{E_\text{ho}}$
\\
\hline
0.1 & 12.329 & 0.010
&2.07(10) & $-18(3)$ & 12.319 & 12.113(10) & 12.011(25)\\
0.08 & 12.287 &  0.018
&2.44(16) & $-16(4)$ & 12.269 & 12.073(13) &12.015(26) \\
0.06 &  12.230 & 0.022
&2.72(25)  &  & 12.208 & 12.045(15)\\
0.05 & 12.204 & 0.025
&2.71(32)  & &   12.179 & 12.043(16)\\
0.04 & 12.184 & 0.035
&2.56(55)  & &   12.149  & 12.047(22)\\
\hline
0 & & & & & 12.015 & 12.019 & \\
\hline
\hline
\end{tabular}
\label{table44}
\end{table*}

\begin{figure}[htbp]
\includegraphics[angle=0,width=84mm]{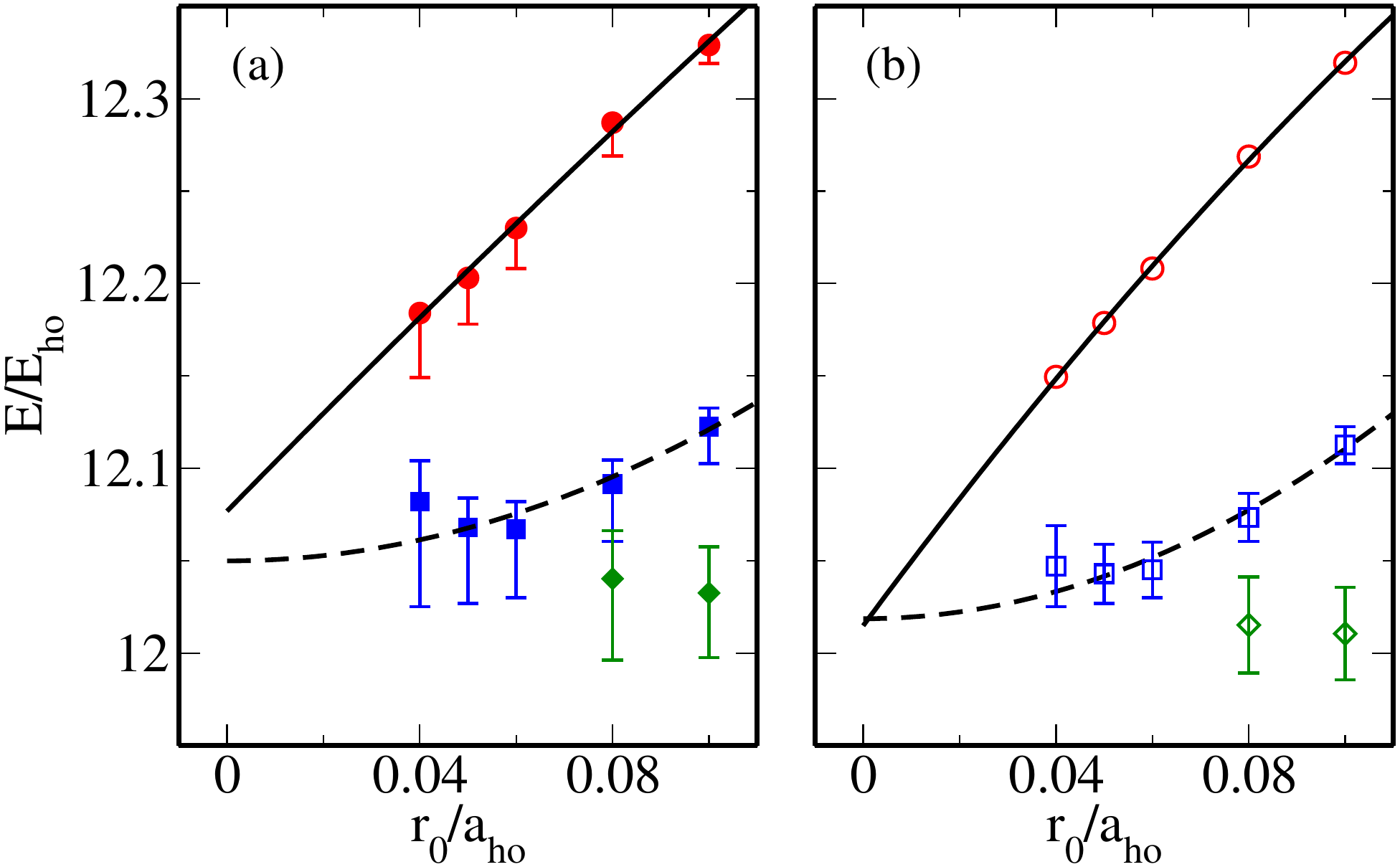}
\caption{(Color online) Ground state energy
of the (4,4) system at unitarity. 
(a) Circles, squares, and diamonds show
$E_{\text{ZRA}, 0}(r_0)$, $E_{\text{ZRA}, 1}(r_0)$, 
and $E_{\text{ZRA}, 2}(r_0)$,
respectively, for the largest basis set considered.
The errorbars of $E_{\text{ZRA}, 0}(r_0)$
show the estimated basis
set error $\Delta E(r_0)$
(see column 3 of Table~\ref{table44});
they extend below the data points but not above.
The errorbars of $E_{\text{ZRA}, 1}(r_0)$
combine the estimated basis set error
and the error of $E^{(1)}(r_0)$
(see column 4 of Table~\ref{table44}).
Lastly, the errorbars of $E_{\text{ZRA}, 2}(r_0)$
combine the estimated basis set error, and
the errors of $E^{(1)}(r_0)$ and
$E^{(2)}(r_0)$
(see column 5 of Table~\ref{table44}).
Solid and dashed lines show the extrapolations
of $E_{\text{ZRA}, 0}(r_0)$ and $E_{\text{ZRA}, 1}(r_0)$
to the zero-range limit.
(b) Same quantities as in (a) but corrected for the estimated
basis set errors. The open symbols show the energies
$E_{\text{ZRA}, 0}(r_0)$, $E_{\text{ZRA}, 1}(r_0)$, 
and $E_{\text{ZRA}, 2}(r_0)$
reported in the last three columns of Table~\ref{table44}.
The uncertainty of $\Delta E(r_0)$ is not accounted for
by the errorbars.
It can be seen that the basis set error lowers the zero-range
energy by about $0.05E_\text{ho}$ or, equivalently,
$0.4\%$.
}
\label{fig_energy44}
\end{figure}

For systems with up to six particles
(see the Supplemental Material~\cite{sm}
for a summary), we believe that
our basis sets for all $r_0$ are very
close to complete.
Specifically, (i) the energy changes very little
upon further enlargement of the basis set,
(ii) the first- and second-order derivatives
$E^{(1)}(r_0)$ and $E^{(2)}(r_0)$ are stable
and their errorbars can be estimated reliably,
(iii) the extrapolations of $E_{\text{ZRA}, 0}(r_0)$, 
$E_{\text{ZRA}, 1}(r_0)$, 
and $E_{\text{ZRA}, 2}(r_0)$
are in very good agreement, and 
(iv) the quantities $3E(r_0)-4V_\text{tr}(r_0)$
and $E_{\text{ZRA}, 1}(r_0)$ agree quite well
[see the discussion after Eq.~(\ref{eq_taylor3})].
For systems with more than six particles,
the construction of a nearly complete basis
set is more challenging, especially for small $r_0$.
As an example, we discuss the (4,4) system;
for the (4,3), (5,4), and (5,5) systems, the reader is
referred to the Supplemental Material~\cite{sm}.

Table~\ref{table44} summarizes our ECG results
for the (4,4) system; the format is the same
as that in Table~\ref{table32} for the (3,2) system.
The smallest range considered
and the errorbars for the $N=8$ system are larger
than those for the $N=5$ system.
For a fixed $r_0$, the quantities $E^{(1)}(r_0)$ and $E^{(2)}(r_0)$
for the $N=8$ system
are about twice as large as for the $N=5$ system and their
errorbars are notably larger. The overall trends, however,
are similar: (i) the energy $E(r_0)$ decreases with decreasing
range, (ii) $E^{(1)}(r_0)$ increases with decreasing range
for $r_0\geq 0.06a_\text{ho}$
(for smaller $r_0$, the trend reverses; we believe
that this is a consequence of the numerics and not
a real trend),
and (iii)
$E^{(2)}(r_0)$ becomes less negative with decreasing range.
Our numerics are not good enough to determine
$E^{(2)}(r_0)$ for $r_0\leq0.06a_\text{ho}$. It can be seen,
however, that the
$E_{\text{ZRA}, 2}(r_0)$ for $r_0=0.1a_\text{ho}$
and $0.08a_\text{ho}$ agree quite well with
the energies obtained by extrapolating 
$E_{\text{ZRA}, 0}(r_0)$ and $E_{\text{ZRA}, 1}(r_0)$
to the zero-range limit (see the last row of Table~\ref{table44}).
Since the basis set error is non-negligible
for $N=8$, Fig.~\ref{fig_energy44} shows
the energies $E_{\text{ZRA}, 0}(r_0)$ (circles), 
$E_{\text{ZRA}, 1}(r_0)$ (squares), 
and $E_{\text{ZRA}, 2}(r_0)$ (diamonds)
before correcting for the basis set error [Fig.~\ref{fig_energy44}(a)]
and after correcting for the basis set error 
[Fig.~\ref{fig_energy44}(b)].
We extrapolate $E_{\text{ZRA}, 0}(r_0)$ and
$E_{\text{ZRA}, 1}(r_0)$ for both the
largest basis set considered [Fig.~\ref{fig_energy44}(a)]
and the
infinite basis set [Fig.~\ref{fig_energy44}(b)]
to the zero-range limit by performing fits of
the form $c_0+c_1 r_0+c_2 r_0^2$
and $c_0+c_2 r_0^2$, respectively
(see solid and dashed lines in Fig.~\ref{fig_energy44}).
To fit a function to $E_{\text{ZRA}, 1}(r_0)$,
the data points are weighted by the inverse square
of the uncertainty.
For the (4,4) and larger
systems, we do not fit to higher-order
polynomials because (i) the number
of data points is five or less and (ii) the errorbars are too
large to determine the $r_0^3$ dependence reliably.
An alternative fit approach that includes the $r_0^3$
term is discussed in Appendix A.

\begin{table*}[htbp]
\caption{
Summary of our zero-range ground state 
energies at unitarity and comparison with literature results
for systems with $N\leq10$ and $N_1-N_2\leq1$.
Column 2 reports the zero-range ground state
energies $E(0)$ calculated
in this work. 
Columns 3 to 16 report
ground state energies from the literature 
calculated by different methods
and their percentage differences from
$E(0)$. 
All energies are reported
in units of $E_\text{ho}$.
The ground state energy of the (2,1) system,
obtained semi-analytically~\cite{werner06}, is
$4.272724E_\text{ho}$.
See text for more details.
}\centering
\begin{tabular}{c c c c c c c c c c c c c c c c}
\hline
\hline
$(N_1,N_2)$ & $E(0)$ 
 & $E_\text{ECG}$     & \%
 & $E_\text{DMC1}$\footnotemark[1]    & \% 
 & $E_\text{DMC2}$\footnotemark[2] & \% 
 & $E_\text{AFMC2}$\footnotemark[3] &  \%  
 & $E_\text{AFMC4}$\footnotemark[3]  &\% 
 & $E_\text{CI}$\footnotemark[4] & \%  
 & $E_\text{lattice}$\footnotemark[5] & \%\\
\hline
(2,1) & 4.2727(1)
&  
&    & 4.281(4) & 0.2 &  
&      &     &    &     &     & 4.279 & 0.1 &  &\\
(2,2) & 5.0091(4)
&  5.0092(4)\footnotemark[6]
&   & 5.051(9)  & 0.8
&  5.028(2) & 0.4  &  &  &  &  & 5.138 & 2.6
&  5.071(+32/$-75$) & 1.2\\
(3,2) &  7.455(1)
&  7.457(3)\footnotemark[7]
&  & 7.61(1)  & 2.1  & 
&      &      &      &    &     &         &          \\
(3,3) &  8.337(4)
& 8.34(9)\footnotemark[7]
&    & 8.64(3)  & 3.6
& 8.377(3)  & 0.5  & 8.26(1) & $-0.9$
& 8.21(1) &  $-1.6$ & 8.601 &  3.2
&  8.347(+80/$-66$)  & 0.1\\
(4,3) &   11.01(2) 
&    &   & 11.36(2) & 3.2    
&     &    &   &   &   &  & 11.021 & 0.1\\
(4,4) &  12.02(3) 
& 12.13(1)\footnotemark[8]
& 0.9  &  12.58(3)  &  4.7  
& 12.04(1) &  0.2  & 11.82(2) &  $-1.7$
&  11.76(3)  &  $-2.2$  & 12.179  & 1.3
& 11.64(+11/$-12$) & $-3.2$\\
(5,4) &  15.24(9)       
&      &     &  15.69(1)  & 3.0
&  &  &  &  &  &  &  &   \\
(5,5) &  16.12(6)        
&      &     &  16.80(4) & 4.2 & 16.10(1) &  $-0.1$
& & & & & &
& 16.05(+3/$-7$) & $-0.4$\\
\hline
\hline
\footnotetext[1]{From Table II of Ref.~\cite{blume07};
 the energies have been calculated for the square well
 potential with range $r_0=0.01a_\text{ho}$,
 corresponding to $r_\text{eff}=0.01a_\text{ho}$.}
\footnotetext[2]{Read off from
Fig. 4 of Ref.~\cite{carlson14}; the FN-DMC
 energies have been extrapolated to the zero-range limit.}
\footnotetext[3]{Read off from
 Fig. 4 of Ref.~\cite{carlson14}; 
 the errorbars only account for the
 statistical uncertainty.
A leading-order correction scheme has been
applied to convert the finite lattice results to the infinite
lattice limit~\cite{carlson14, carlson}.}
\footnotetext[4]{From Table I of Ref.~\cite{alhassid13};
 the CI energies have been obtained for a 
finite shell model space and
the two-body coupling constant has been renormalized by
matching the two-particle ground state energy
to the exact energy.} 
\footnotetext[5]{From Table VI of Ref.~\cite{kaplan11};
the upper and lower limits of the errorbars are different
and separated by a slash. The errorbars account for
statistical, fitting, finite volume,
and spatial discretization errors, but do not
account for systematic errors due to the
contributions from excited states.
We note that
odd $N$ systems were considered in Ref.~\cite{kaplan10}.
The results in Ref.~\cite{kaplan10}
were described as ``preliminary" and are
not included here.
}
\footnotetext[6]{From Ref.~\cite{javier09};
the energy has been obtained
by solving the hyperangular Schr\"odinger
equation.}
\footnotetext[7]{From Table XXI of Ref.~\cite{cg_rmp}; 
the energies have been extrapolated to 
the zero-range limit using the zeroth-order 
extrapolation scheme.}  
\footnotetext[8]{From Table II of Ref.~\cite{quiney14};
the energy 
has been extrapolated to the zero-range limit
using the zeroth-order extrapolation scheme.}
\end{tabular}
\label{table1}
\end{table*}

Table~\ref{table1} summarizes our
zero-range ground state energies $E(0)$ (column 2) 
obtained by
extrapolating the energies $E_\text{ZRA,1}(r_0)$,
which have been shifted down by the estimated
basis set
error, to the zero-range limit.
The errorbars given in parenthesis are
estimated by combining the zero-range
extrapolation error, the uncertainty of the basis
set error, and the uncertainty of $E^{(1)}(r_0)$.

Our ground state energy for the (2,1) system
agrees excellently with the semi-analytic energy obtained
using the zero-range framework of Ref.~\citep{werner06}.
For comparison, Table~\ref{table1} 
includes the ground state
energies from the
literature obtained by various methods.
Column 3 of Table~\ref{table1} reports the
zero-range ground state energies $E_\text{ECG}$
calculated by the
ECG method in previous works~\cite{javier09, 
cg_rmp, quiney14}. 
Our results for the (2,2), (3,2) and (3,3) systems
agree within errorbars with the literature results.
For the (3,3) system, we provide a notably tighter
errorbar. The ECG energy 
for the (4,4) system by Bradly {\em{et al.}}~\cite{quiney14}
is about $0.9\%$ higher than our (4,4) energy.
Bradly {\em{et al.}} estimate that the error due to the use
of a restricted basis set is about $0.6\%$
for the relative energy, translating to $0.53\%$
for the total energy.
This estimate is in reasonable agreement with 
the difference between
their energy and our energy.
Column 5 reports the FN-DMC energies
$E_\text{DMC1}$ calculated for the square well potential with range 
$0.01a_\text{ho}$~\cite{blume07}.
The deviations between the FN-DMC and our ECG energies
are due to the positive effective range correction 
and the approximate nature of the nodal surface of the trial
wave function (the latter dominates). Column 7 reports
highly-improved FN-DMC energies $E_\text{DMC2}$~\cite{carlson14}.
These energies have been extrapolated to the zero-range limit.
The FN-DMC energies from Ref.~\cite{carlson14}
agree very well with our ECG energies (the agreement
is better than $0.6\%$ and, for $N=8$ and $10$, the errorbars
overlap).
Unfortunately, Ref.~\cite{carlson14} considered
only spin-balanced systems.
Columns 9 and 11 report
the energies $E_\text{AFMC2}$
and $E_\text{AFMC4}$ calculated using the
AFMC approach with $q^2$ and $q^2+q^4$
dispersion relations, respectively~\citep{carlson14}.
These energies
have been obtained by applying a leading-order
correction scheme to convert the finite lattice results to
the infinite lattice limit but
have not been extrapolated
to the infinite lattice size limit~\cite{carlson14, carlson}.
Note that the AFMC energies for fixed $N$
but different dispersion relations do not
agree within errorbars. The reason may be
that the corrections due to the finite
lattice spacing behave differently for the
different dispersion relations
and that the errorbars are purely statistical. 
Column 13 reports the configuration 
interaction (CI) energies $E_\text{CI}$
obtained using a limited CI shell model space~\cite{alhassid13}.
The authors of Ref.~\cite{alhassid13} noted that the two-body
interaction strength was renormalized using
an approach that could be improved upon.
Improvement to both these aspects
(enlarged CI model space and refined renormalization
approach)
 could change the CI
energies.
Interestingly,
the odd $N$ CI energies agree quite well
with our ECG energies while the even $N$ CI energies
are higher by between $1.3\%$ and $3.2\%$.
It is not clear to us what the origin of the different
even and odd $N$ behaviors is.
Column 15 reports the lattice MC
energies $E_\text{lattice}$~\cite{kaplan11}.
The lattice MC energies exhibit
shell effects that are absent in the FN-DMC,
AFMC, and---for small $N$---ECG energies
(our energies for $N\leq10$ do not
exhibit shell effects). 
The lattice MC energy for $N=8$
is $3.2\%$ lower
than our ECG, reflecting
the shell effects exhibited by the lattice MC
energies in the small $N$ regime.
For systems with $N>10$, the lattice
MC energies are higher than or equal to (within errorbars) the
FN-DMC energies $E_\text{DMC2}$ from Ref.~\cite{carlson14}.
The difference between the lattice MC
energies and FN-DMC energies for $N>10$
is smallest for closed shell systems.
Besides the results summarized in
Table~\ref{table1}, we also compared our ground
state energies 
with DFT energies~\cite{bulgac07}
for both even and odd $N$ systems.
The DFT energies are $5\%$ to $10\%$ higher than our ECG energies.

\begin{table}[htbp]
\caption{Zero-range
contact $C(0)$ at unitarity for $N=3-10$.
Column 2 reports the zero-range
contact $C(0)$ determined using the
adiabatic energy relation. 
$C(0)$ for the $(1,1)$ system, obtained
analytically 
from the implicit eigenequation derived
in Ref.~\cite{busch98}, is 
$4\sqrt{2\pi}a_\text{ho}^{-1}
=10.026513a_\text{ho}^{-1}$.
$C(0)$ for the $(2,1)$ system, obtained
semi-analytically using the 
hyperspherical coordinate framework~\cite{werner06, 
greene10, yan13}, is 
$10.468967a_\text{ho}^{-1}$.
}\centering
\begin{tabular}{c c}
\hline
\hline
$(N_1, N_2)$ & $C(0) a_\text{ho}$ \\
\hline
(2,1) & 10.469(1) \\
(2,2) & 25.74(1) \\
(3,2) & 25.20(1) \\
(3,3) & 40.39(8)  \\
(3,4) & 38.2(2)  \\
(4,4) & 55.4(5)  \\
(5,4) & 56.9(9) \\
(5,5) & 72.3(8) \\
\hline
\hline
\end{tabular}
\label{table4}
\end{table}

\begin{figure}[htbp]
\includegraphics[angle=0,width=80mm]{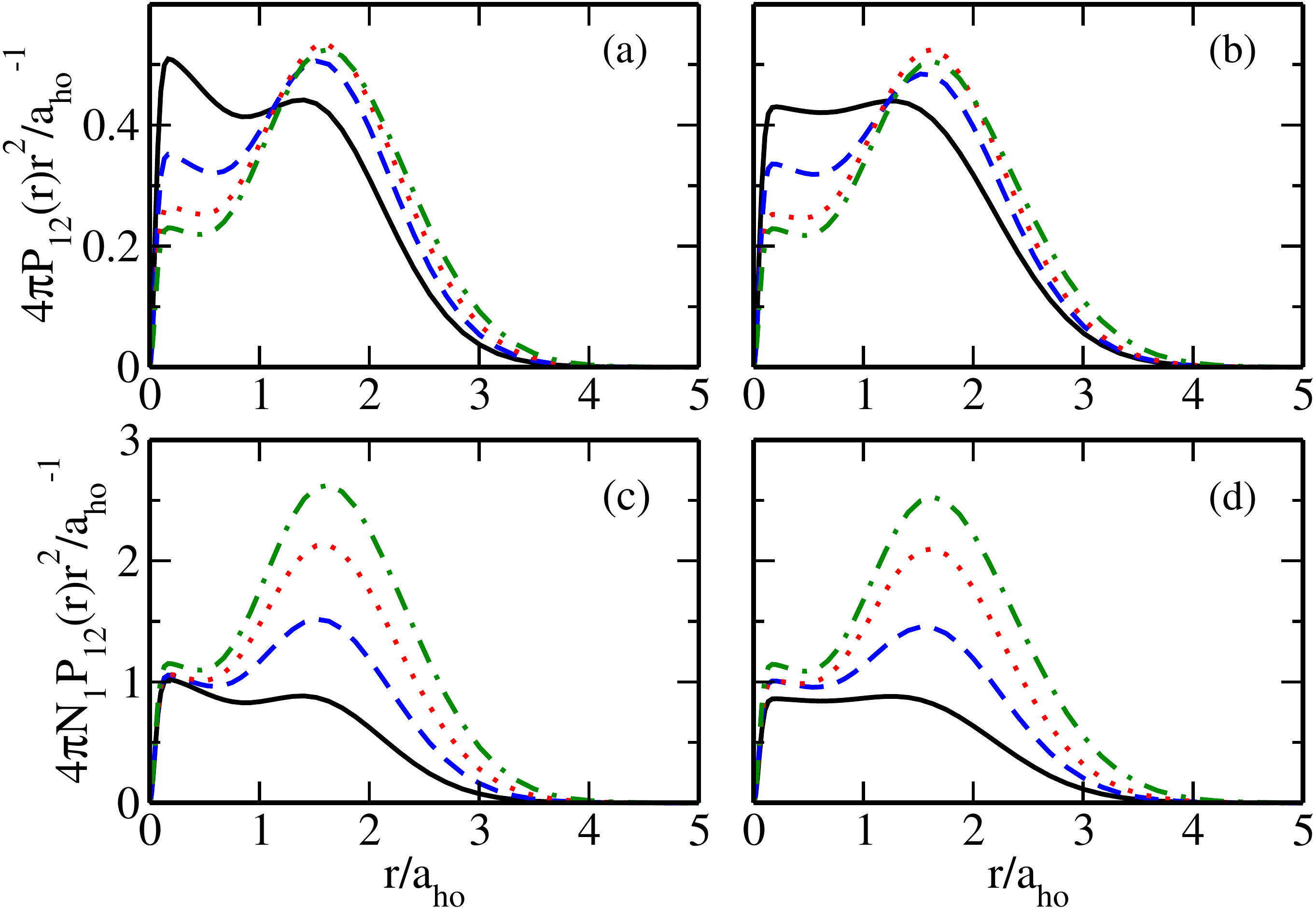}
\caption{(Color online) Panels (a) and (c) show the
scaled pair distribution functions
$4\pi P_\text{12}(r)r^2$
and $4\pi N_1 P_\text{12}(r)r^2$,
respectively,
of the ground state at unitarity for
the (2,2) system (solid line), (3,3) system (dashed line),
(4,4) system (dotted line), and (5,5) system (dash-dotted line).
Panels (b) and (d) show the
scaled pair distribution functions
$4\pi P_\text{12}(r)r^2$ 
and $4\pi N_1 P_\text{12}(r)r^2$,
respectively,
of the ground state at unitarity for
the (2,1) system (solid line), (3,2) system (dashed line),
(4,3) system (dotted line), and (5,4) system (dash-dotted line).
The calculations
are performed for $r_0=0.06a_\text{ho}$.}
\label{fig_pair}
\end{figure}

In addition to the energies, we calculate the contact at unitarity.
To remove the leading-order range dependence, we analyze
the quantities $C_{\text{ZRA},0}(r_0)$ and
$C_{\text{ZRA},1}(r_0)$. While the energies
$E_{\text{ZRA},0}(r_0)$ and
$E_{\text{ZRA},1}(r_0)$ approach the $r_0=0$ limit from
above and below, respectively, for all $N$ considered,
the  contacts $C_{\text{ZRA},0}(r_0)$ and
$C_{\text{ZRA},1}(r_0)$ approach the $r_0=0$
limit from either above or below.
Specifically, fitting $C_{\text{ZRA},0}(r_0)$ to
a function of the form $c_0+c_1 r_0+c_2 r_0^2$,
we find that $c_1$ is positive for $N=4$,
very close to zero for $N=6$, negative for $N=8$,
and again positive for $N=10$.
For the odd $N$ systems, $c_1$ is always positive.
The pair distribution functions exhibit an analogous
range-dependence in the $r_0\ll r\ll a_\text{ho}$
region (see Figs. 9 and 10 of the Supplemental Material
for the $N=5$ and 8 systems),
suggesting that the intricate $N$ and $r_0$ dependence of
the contact is a real effect and not an artifact
of our numerics. Our convergence studies
support this interpretation.
Table~\ref{table4} reports the zero-range contact
$C(0)$ for $N=3-10$ at unitarity obtained by extrapolating
$C_{\text{ZRA},1}(r_0)$ to the zero-range limit.
The errorbars in parenthesis account for the
zero-range extrapolation error and the basis set error.
The $r_0=0$ extrapolations of $C_{\text{ZRA},0}(r_0)$
and of the contact extracted from the pair distribution
functions agree with the values reported in Table~\ref{table4}
but have larger errorbars.
The contact exhibits an interesting even-odd pattern.
Specifically, for the $N=4$ and 5 systems 
(and the 6 and 7 systems, and the 8 and 9 systems),
the contacts are roughly equal, reflecting the
fact that these neighboring even-odd systems contain
the same number of pairs.
To zeroth-order, the contact scales as $N_2$ times
the contact of the two-body system,
i.e., linearly with the number of pairs. 
Since $C(0)$ scales with $N_2$,
the $r_0\ll r\ll a_\text{ho}$ region
of the scaled pair distribution functions 
$4\pi P_{12}(r) r^2$ approximately collapse to a single
curve if multiplied by $N_1$. This approximate
collapse is illustrated in Figs.~\ref{fig_pair}(c) and
\ref{fig_pair}(d).

\begin{figure}[htbp]
\includegraphics[angle=0,width=60mm]{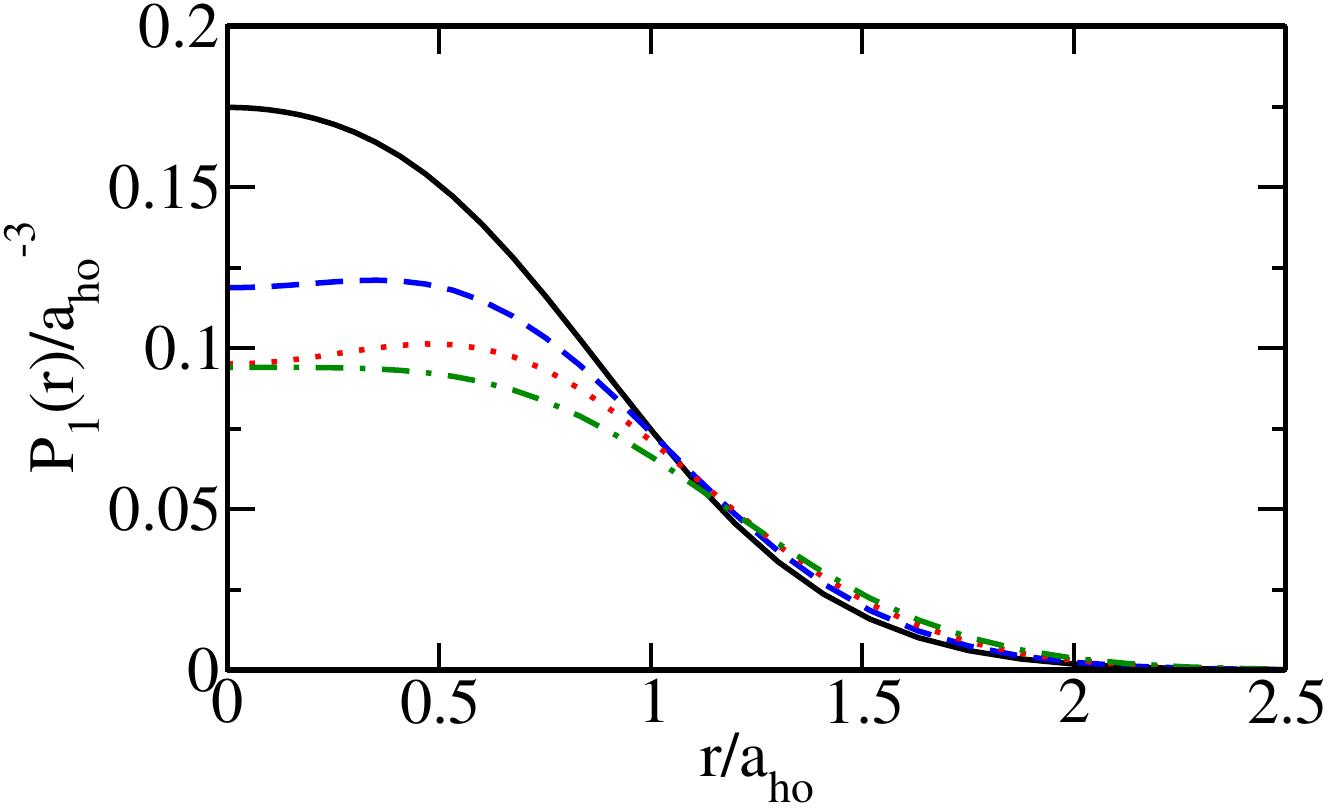}
\caption{(Color online) Radial density
$P_\text{1}(r)$ of the ground state at unitarity for
the (2,2) system (solid line), (3,3) system (dashed line),
(4,4) system (dotted line), and (5,5) system (dash-dotted line).
The calculations
are performed for $r_0=0.06a_\text{ho}$.}
\label{fig_reven}
\end{figure}

\begin{figure}[htbp]
\includegraphics[angle=0,width=60mm]{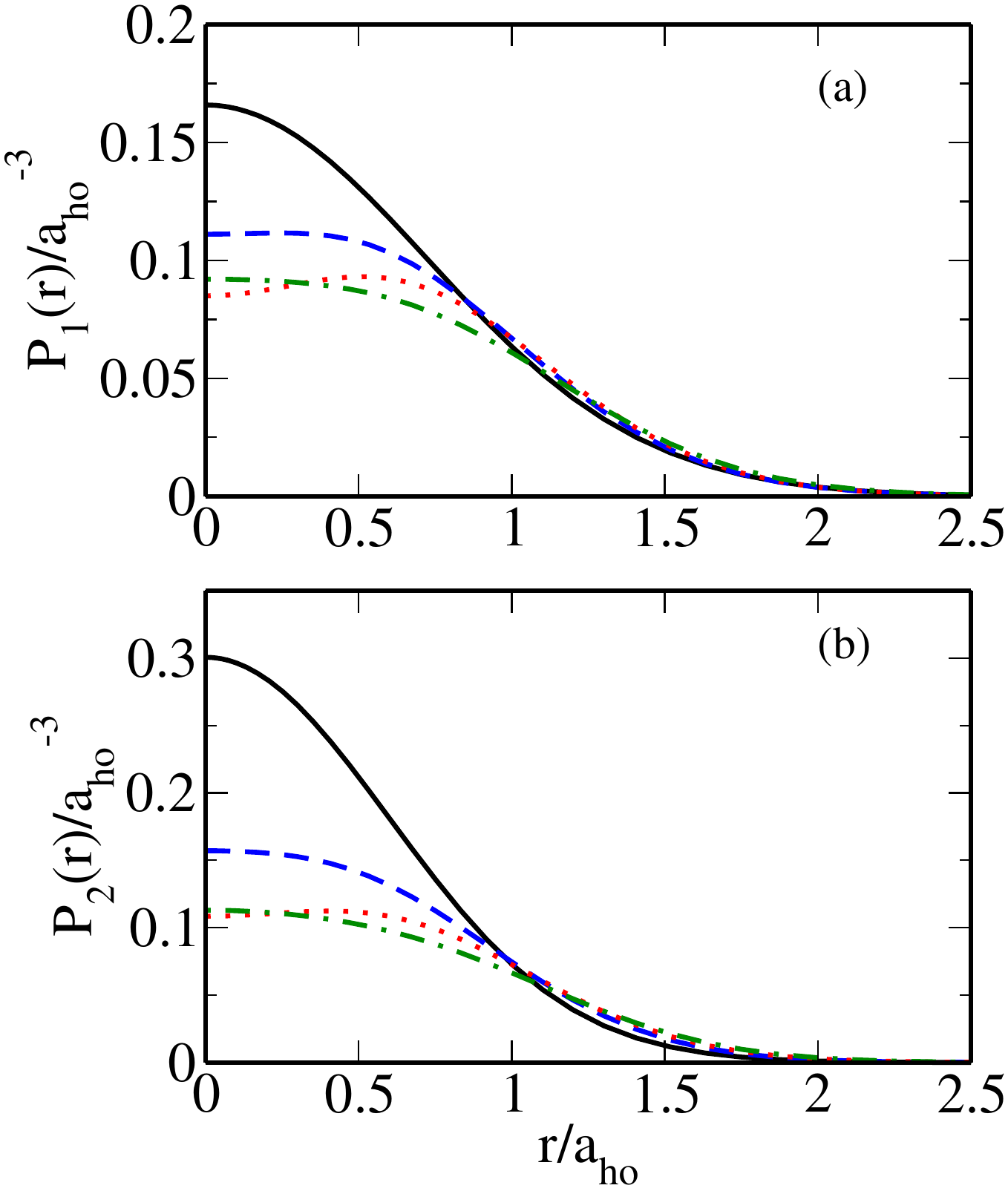}
\caption{(Color online) Panels (a) and (b) show
the
radial density of the majority species
$P_1(r)$ and the minority species $P_2(r)$, respectively,
for the ground state of
the (2,1) system (solid line), (3,2) system (dashed line),
(4,3) system (dotted line), and (5,4) system (dash-dotted line).
The calculations
are performed for $r_0=0.06a_\text{ho}$.}
\label{fig_rodd}
\end{figure}

Figure~\ref{fig_reven} shows the
radial density
$P_\text{1}(r)$ of the ground state at unitarity
for even $N$ and $r_0=0.06a_\text{ho}$.
We note that the convergence of the radial density
in the small $r$ regime is not as good as
that of the pair distribution function,
especially for large $N$ and small $r_0$.
$P_\text{1}(r)$ peaks at  $r=0$
for the (2,2) system, is relatively flat
in the small $r$ region for the (3,3) and (5,5) systems,
and peaks around $0.6a_\text{ho}$
for the (4,4) system.
To estimate the range dependence,
we calculate $P_\text{1}(r)$ for different $r_0$
for the (2,2), (3,3), and (4,4) systems.
For a given system, 
the $r\lesssim 0.5a_\text{ho}$ region of
$P_\text{1}(r)$ increases
with decreasing
two-body range $r_0$ (see 
Fig. 8 of the Supplemental Material~\cite{sm}).
The changes with $r_0$ are relatively small and the
densities displayed in Fig.~\ref{fig_reven}
show the generic behavior of trapped Fermi gases
with short-range interactions.
Figure~\ref{fig_rodd} shows
$P_{j}(r)$, $j=1$ and $2$,
for the odd $N$ systems
at unitarity for $r_0=0.06a_\text{ho}$.
$P_\text{1}(r)$ and $P_\text{2}(r)$
peak at  $r=0$
for the (2,1) system,
are relatively flat
in the small $r$ region for the (3,2) and (5,4) systems,
and peak around $0.5a_\text{ho}$
for the (4,3) system.
We  find that the range-dependence
of the radial density for the odd
$N$ systems is similar to that for the even $N$ systems
(see Fig. 7 of the Supplemental Material~\cite{sm}).

\begin{figure}[htbp]
\includegraphics[angle=0,width=60mm]{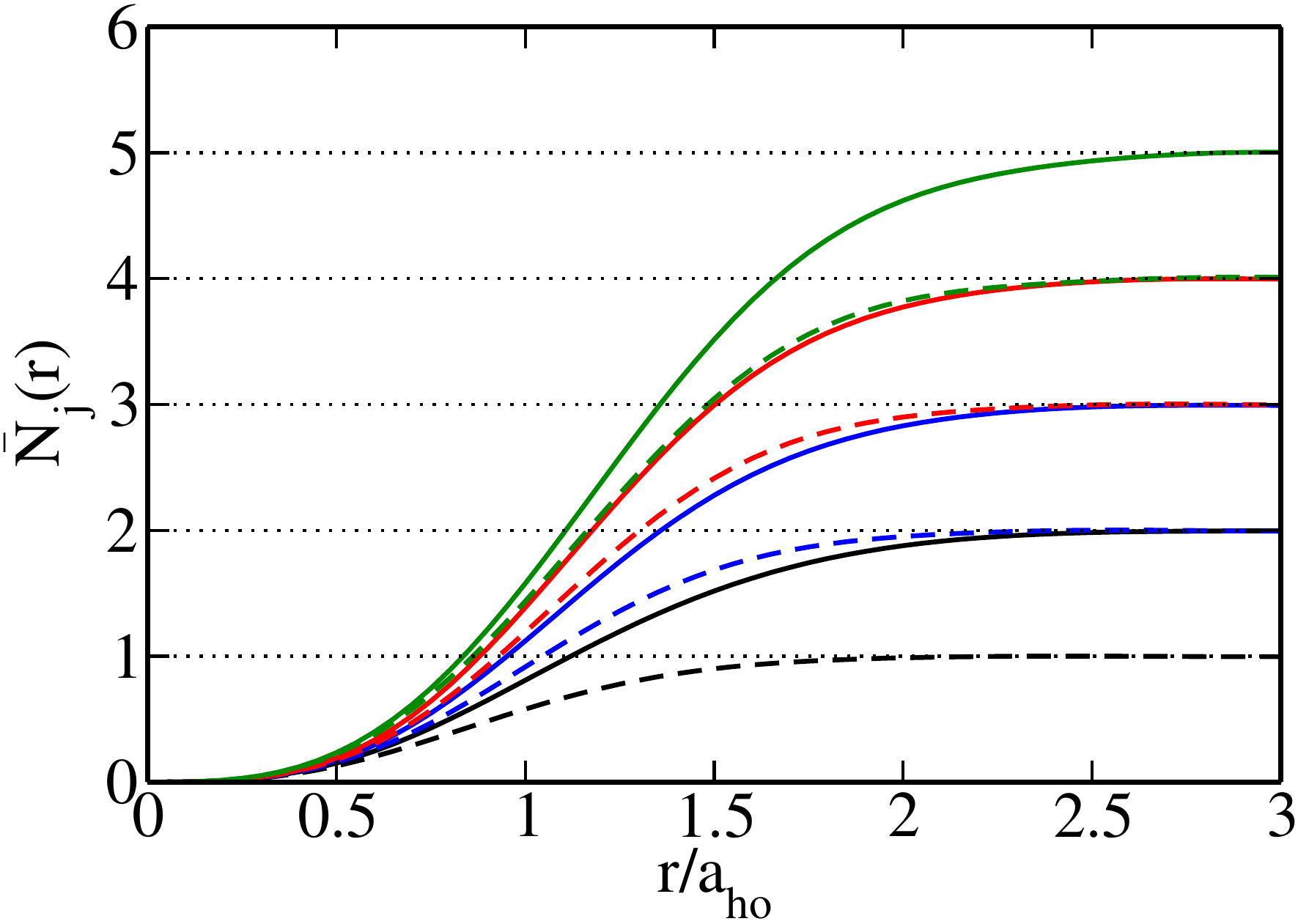}
\caption{(Color online) Solid and dashed
lines show the integrated quantities $\bar{N}_1(r)$
and $\bar{N}_2(r)$, respectively,
for odd $N$ systems as a function
of $r$. From bottom to top,
the curves correspond to systems with
$N=3$, 5, 7, and 9.
The horizontal dotted lines at 1 to 5 serve as a
guide to the eye.
The calculations
are performed for $r_0=0.06a_\text{ho}$.}
\label{fig_radial}
\end{figure}

To gain insights into the pairing of the particles,
Fig.~\ref{fig_radial} shows the integrated quantities
$\bar{N}_j(r)$,
\begin{eqnarray}
\bar{N}_j(r)=4\pi N_j\int_{0}^{r}P_{j}(r')r'^2dr',
\end{eqnarray} 
for the odd $N$ systems.
Solid and dashed lines show $\bar{N}_j(r)$
for the majority ($j=1$)
and minority ($j=2$) species, respectively.
$\bar{N}_j(r)$ monitors the number of particles
of species $j$ located
between zero and $r$, and approaches
$N_j$ in the large $r$ limit. 
We find that 
$\bar{N}_1(r)$ and $\bar{N}_2(r)$ take,
for $N$ fixed, different values for all $r$, suggesting
that there
exists no core region where the systems are fully paired.
This is in contrast to an earlier FN-DMC study~\cite{blume08},
which suggested that the $N=9$ system has a fully
paired core. It should be noted that a fully
paired core is expected in the large $N$ limit~\cite{son07};
however, how many particles are needed to be in the large
$N$ limit is not clear.

\section{Conclusions}
\label{sec_conclusion}

This paper considered the ground state properties
of trapped two-component Fermi gases at unitarity
with up to ten particles.
The calculations were performed for interspecies finite-range
Gaussian interaction potentials using the ECG approach.
Previous ECG calculations were limited
to $N=3-6$ and 8. The present work additionally
considered the spin-imbalanced $N=7$ and 9 systems
with $L^{\pi}=1^{-}$ symmetry and the spin-balanced 
$N=10$ system with $L^{\pi}=0^{+}$ symmetry.
A new range-correction scheme, which allows for the
leading and---in some cases---the sub-leading range
dependence to be removed, was introduced.
The accuracy of the range correction scheme was tested 
extensively for small $N$ systems ($N\leq6$)
and then applied to larger systems ($N=7-10$).
The resulting extrapolated zero-range energies
have errorbars that range from $0.002\%$
for $N=3$ to $0.6\%$ and $0.4\%$ for $N=9$
and 10. The energies agree well with the
FN-DMC energies from  Ref.~\cite{carlson14},
suggesting that the zero-range energies of harmonically
trapped two-component Fermi gases with
$N\leq10 $ ($N_1-N_2=0$ or 1) are now known
with an accuracy better than $1\%$. The finite-range
energies were reported for finite $r_0$ and all $N$.
These finite-range energies provide variational
upper bounds and are expected to help assess
the accuracy of future finite-range calculations
(the range $r_0$ can be easily converted to the
effective range).
In addition to the energy, the pair distribution
functions and radial densities were analyzed.
The Tan contacts obtained through the
adiabatic and pair relations were found to agree within
errorbars.

\section{Acknowledgements}

We gratefully acknowledge discussions
with Y. Yan, Y. Alhassid, and M. M. Forbes,
email correspondence with J. Carlson,
and support by the National
Science Foundation (NSF) through Grant No. PHY-1205443.
This work used the Extreme Science and Engineering
Discovery Environment (XSEDE), 
which is supported by NSF Grant No. ACI-1053575,
and the WSU HPC.

\appendix
\section{Additional
comments on the range-correction scheme}

In the main text, we independently fit the quantities
$E_{\text{ZRA},0}(r_0)$, $E_{\text{ZRA},1}(r_0)$,
and $E_{\text{ZRA},2}(r_0)$. The resulting zero-range
energies were found to be in good agreement.
This appendix discusses that a single correlated
fit yields results that are consistent with those 
obtained from the independent fits.

We assume that the ground state energy 
$E(r_0)=E_\text{ZRA,0}(r_0)$
is a polynomial in the two-body interaction range
$r_0$,
\begin{eqnarray}
\label{eq_a1}
E(r_0)=c_0+c_1 r_0+c_2 r_0^2 +c_3 r_0^3
+\mathcal{O}(r_0^4).
\end{eqnarray}
Using Eq.~(\ref{eq_a1}) to calculate
$E^{(1)}(r_0)$ and $E^{(2)}(r_0)$ and inserting
the results into
Eq.~(\ref{eq_taylor3}), we find
\begin{eqnarray}
\label{eq_a2}
E_\text{ZRA,1}(r_0)=c_0-c_2r_0^2-2c_3r_0^3+\mathcal{O}(r_0^4)
\end{eqnarray}
and
\begin{eqnarray}
\label{eq_a3}
E_\text{ZRA,2}(r_0)=c_0+c_3r_0^3+\mathcal{O}(r_0^4).
\end{eqnarray}
As expected, the leading-order range-dependencies
of  $E_\text{ZRA,1}(r_0)$ and $E_\text{ZRA,2}(r_0)$ are
quadratic and cubic, respectively,
and the functional forms of $E_{\text{ZRA},j}(r_0)$
are not independent.
Specifically, the quadratic coefficient of $E_\text{ZRA,1}(r_0)$
has the opposite
sign but the same magnitude as
that of $E_\text{ZRA,0}(r_0)$,
the cubic coefficient of $E_\text{ZRA,1}(r_0)$
has the opposite sign but twice
the magnitude as that of $E_\text{ZRA,0}(r_0)$,
and the cubic coefficient of $E_\text{ZRA,2}(r_0)$
is the same as that of $E_\text{ZRA,0}(r_0)$.
Interestingly, our independent fits shown in
Figs.~\ref{fig_energy32} and \ref{fig_energy44}
of the main text and Figs.~1-6 of the Supplemental
Material are largely consistent with 
Eqs.~(\ref{eq_a1})-(\ref{eq_a3}).
For example, our fits of $E_\text{ZRA,0}(r_0)$
yield a negative $r_0^2$ coefficient and those of
$E_\text{ZRA,1}(r_0)$ yield a positive
$r_0^2$ coefficient.
The magnitudes of these coefficients, however,
depend fairly sensitively on the number of
terms included in the independent fits. 

As an alternative, we perform a simultaneous 
four-parameter fit of 
$E_{\text{ZRA},0}(r_0)$, $E_{\text{ZRA},1}(r_0)$,
and $E_{\text{ZRA},2}(r_0)$
using Eqs.~(\ref{eq_a1})-(\ref{eq_a3}).
Each data point is weighted by the inverse square
of the uncertainty [for $E_\text{ZRA,0}(r_0)$
we assume an uncertainty of $0.3\Delta E(r_0)$
and the uncertainties of $E_\text{ZRA,1}(r_0)$
and $E_\text{ZRA,2}(r_0)$ are given in Tables~\ref{table32}
and \ref{table44} of the main text
and Tables I-VI of the Supplemental Material].
The resulting zero-range energies for 
$N=3-10$
are $4.2726E_\text{ho}$, 
$5.0088E_\text{ho}$, $7.454E_\text{ho}$, 
$8.335E_\text{ho}$, $11.01E_\text{ho}$, 
$12.02E_\text{ho}$, $15.25E_\text{ho}$,
and $16.12E_\text{ho}$, respectively.
These energies lie within the errorbars of the zero-range
energies reported in Table~\ref{table1}.
The simultaneous fit yields
a positive $c_1$ coefficient and negative
$c_2$ and $c_3$ coefficients for all $N$.
The $c_1$ coefficient obtained from
the independent fit of $E_\text{ZRA,0}(r_0)$
differs from that obtained from the simultaneous fit
by less than $10\%$ for all $N$.

We also apply the simultaneous fit approach
to the contact. We fit our numerically
obtained $C_{\text{ZRA},0}(r_0)$
and $C_{\text{ZRA},1}(r_0)$ to functions
of the form 
$c_0+c_1 r_0+c_2 r_0^2+c_3 r_0^3$ and
$c_0-c_2r_0^2-2c_3r_0^3$, respectively,
for $N<6$, and to functions
of the form
$c_0+c_1 r_0+c_2 r_0^2$ and
$c_0-c_2r_0^2$, respectively, for 
$N=7-10$.
The resulting zero-range contacts $C(0)$
for $N=3-10$ lie within the errorbars of the
zero-range contacts reported in Table~\ref{table4}.

\bibliographystyle{apsrev4-1}
\nocite{apsrev41Control}
\bibliography{mybib}

\end{document}